\documentclass[]{spie}  %>>> use for US letter paper
\usepackage[]{graphicx}
\usepackage[colorinlistoftodos]{todonotes}
\usepackage{amsmath}

\title{A LEKID-based CMB instrument design for large-scale observations in Greenland} 

\author{D. C. Araujo\supit{a}, P. A. R. Ade\supit{b}, J. R. Bond\supit{c}, K. J. Bradford\supit{d}, D. Chapman\supit{a}, G. Che\supit{d}, \\P. K. Day\supit{e}, J. Didier\supit{a}, S. Doyle\supit{b}, H. K. Eriksen\supit{f}, D. Flanigan\supit{a}, C. E. Groppi\supit{d}, \\S. N. Hillbrand\supit{a,g}, B. R. Johnson\supit{a}, G. Jones\supit{a}, M. Limon\supit{a}, A. D. Miller\supit{a}, P. Mauskopf\supit{d}, \\H. McCarrick\supit{a}, T. Mroczkowski\supit{h}, B. Reichborn-Kjennerud\supit{a}, B. Smiley\supit{a}, J. Sobrin\supit{a}, \\I. K. Wehus\supit{e}, J. Zmuidzinas\supit{e,i}
\skiplinehalf
\supit{a}Columbia University Department of Physics, New York, New York, United States; \\
\supit{b}School of Physics and Astronomy, Cardiff University, United Kingdom;\\
\supit{c}Canadian Institute for Theoretical Astrophysics, University of Toronto, Canada;\\
\supit{d}Department of Physics, Arizona State University, Phoenix, Arizona, United States;\\
\supit{e}Jet Propulsion Laboratory, California Institute of Technology, Pasadena, California, United States;\\
\supit{f}Institute of Theoretical Astrophysics, University of Oslo, Oslo, Norway;\\
\supit{g}Department of Physics and Astronomy, California State University, \\Sacramento, California, United States;\\
\supit{h}U.S. Naval Research Laboratory, Washington, DC, United States;\\
\supit{i}Department of Physics, California Institute of Technology, Pasadena, California, United States
}

\authorinfo{Further author information: send correspondence to Derek C. Araujo. \\E-mail: derek@phys.columbia.edu, Telephone: 1 212 851 9380. 
\\Copyright 2014 Society of Photo-Optical Instrumentation Engineers.  One print or electronic copy may be made for personal use only.  Systematic reproduction and distribution, duplication of any material in this paper for a fee or for commercial purposes, or modification of the content of the paper are prohibited.}
\begin{document} 
\maketitle 

%%%%%%%%%%%%%%%%%%%%%%%%%%%%%%%%%%%%%%%%%%%%%%%%%%%%%%%%%%%%% 

\begin{abstract}
We present the results of a feasibility study, which examined deployment of a ground-based millimeter-wave polarimeter, tailored for observing the cosmic microwave background (CMB), to Isi Station in Greenland.  
The instrument for this study is based on lumped-element kinetic inductance detectors (LEKIDs) and an F/2.4 catoptric, crossed-Dragone telescope with a 500 mm aperture.
The telescope is mounted inside the receiver and cooled to $<\,4$ K by a closed-cycle $^4$He refrigerator to reduce background loading on the detectors. 
Linearly polarized signals from the sky are modulated with a metal-mesh half-wave plate that is rotated at the aperture stop of the telescope with a hollow-shaft motor based on a superconducting magnetic bearing. 
The modular detector array design includes at least 2300 LEKIDs, and it can be configured for spectral bands centered on 150~GHz or greater.  Our study considered configurations for observing in spectral bands centered on 150, 210 and 267~GHz.
The entire polarimeter is mounted on a commercial precision rotary air bearing, which allows fast azimuth scan speeds with negligible vibration and mechanical wear over time.
A slip ring provides power to the instrument, enabling circular scans (360 degrees of continuous rotation). 
This mount, when combined with sky rotation and the latitude of the observation site, produces a hypotrochoid scan pattern, which yields excellent cross-linking and enables 34\% of the sky to be observed using a range of constant elevation scans.
This scan pattern and sky coverage combined with the beam size (15~arcmin at 150~GHz) makes the instrument sensitive to $5 < \ell < 1000$ in the angular power spectra.
\end{abstract}


%%%%%%%%%%%%%%%%%%%%%%%%%%%%%%%%%%%%%%%%%%%%%%%%%%%%%%%%%%%%%

\section{INTRODUCTION}
\label{sec:intro} 

The CMB is a bath of primordial photons that permeates all of space and carries an image of the Universe as it was 380,000 years after the Big Bang.
Measurements of the angular temperature and polarization anisotropies in the CMB have revealed a wealth of information about the origin, composition, large-scale structure and dynamics of the Universe~\cite{planck,wmap,quiet,brown,bischoff,chiang}.
For example, CMB measurements to date have helped reveal that spacetime is flat, the Universe is 14 billion years old, and it is dominated by cold dark matter and dark energy.
These results have played an important role in establishing the standard $\Lambda$CDM cosmological model~\cite{planck}. 
A faint divergence-free ``B-mode'' polarization anisotropy signal in the CMB should contain additional information, and CMB research is now focused on measuring these B-mode signals.
B-modes generated when CMB photons were gravitationally lensed by large-scale structure constrain the values of physical parameters such as the sum of the neutrino masses, while an inflationary gravity-wave (IGW) B-mode signal would reveal the energy scale at which inflation occurred\cite{baumann2009}.
Experiments are starting to detect B-mode polarization in small patches of the Southern sky~\cite{bicep2,hanson,polarbear1,polarbear2}.  Follow-up precision measurements of the entire sky are therefore warranted.

%-------------
   \begin{figure}[t]
   %\centering
\begin{minipage}[c]{0.35\textwidth}
\includegraphics[height=7.5cm]{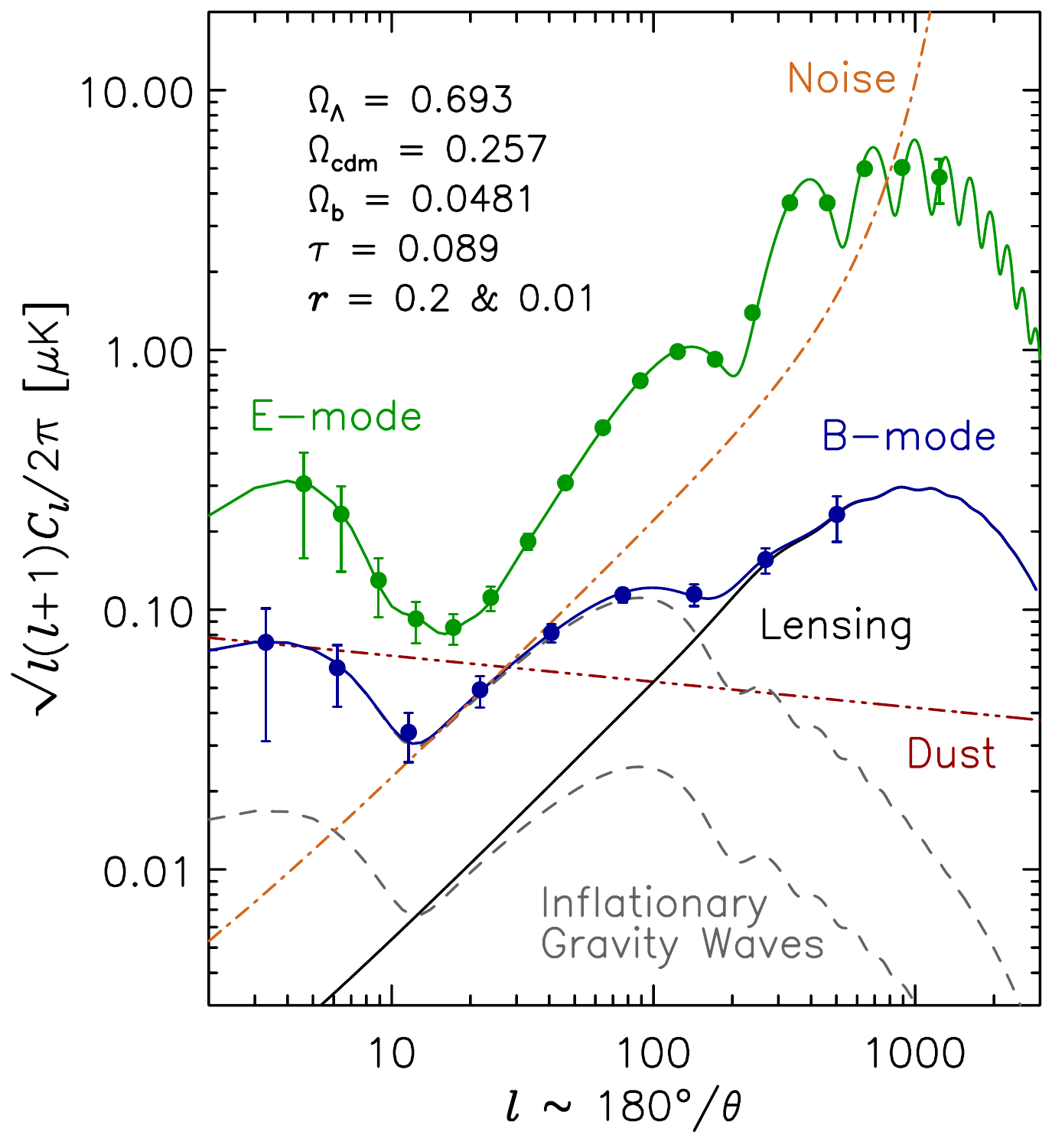}
\end{minipage}
\hspace{1.2cm}
\begin{minipage}[c]{0.55\textwidth}
%\centering
%\begin{tiny}
   \begin{center}
\resizebox{\columnwidth}{!}{%
\begin{tabular}{|l|c|c|}
\hline
& Config. 1 & Config. 2 \\
\hline
Spectral Band Centers [GHz] & 150 & 210, 267 \\
Spectral Bandwidth [$\delta\nu$/$\nu$]  & 0.27   & 0.14, 0.22 \\
Number of Detectors & 2317 & 938, 2345 \\
Total Number of Detectors & 2317 & 3283 \\
Detector NET [$\mu$K$\sqrt{\mbox{sec}}$] & 418 & 1040, 1480 \\
Instrument NET [$\mu$K$\sqrt{\mbox{sec}}$] & 8.68 & 34.0, 30.6  \\
Aperture Diameter [mm] & 500 & 500   \\
Beam FWHM [arcmin] & 15. & 11., 8.4 \\
Total Sky Coverage [deg$^2$] & 14,000 & 14,000 \\
$\ell$ Range & 5 to 1000  & 5 to 1000  \\
\hline
\end{tabular}
}
\end{center}
%\end{tiny}
\end{minipage}
\quad
\vspace{0.25cm}
   \caption[observatory] 
%>>>> use \label inside caption to get Fig. number with \ref{}
   { \label{fig:spectra} 
\textbf{Left:} Science goals.  Shown are theoretical angular power spectra for the E-mode signal, the B-mode signal, noise, and Galactic dust.  The Galactic dust curve was computed assuming a fiducial polarization fraction of 15\%.  The points show the raw sensitivity of the instrument performance for one year of integration time.  The gray dashed lines correspond to anticipated primordial B-mode signals for values of $r$ = 0.2 and 0.01.  To extract foreground contamination, our study includes an instrument configuration tailored for high frequency observations of Galactic dust emission. \textbf{Right:} Projected performance characteristics for the two instrument configurations considered in our study. The 150 GHz NET values were computed assuming a typical loading of 3~pW and a total NEP of 4.9$\times$10$^{-17}$~W/$\sqrt{\mbox{Hz}}$. This assumption is supported by detector noise measurements (see Fig.~\ref{fig:nep}). Our collaboration is currently developing a device in which each pixel detects dual-polarization.  This would double the number of detectors in the focal plane and decrease the instrument NET by a factor of $\sqrt{2}$.}
\vspace{-0.25cm}
   \end{figure} 
   
% ------------------------------ 

In this paper, we describe the design of a compact CMB polarimeter based on lumped-element kinetic inductance detectors (LEKID)s.
This example instrument is designed with the sky coverage, angular resolution, sensitivity, and spectral coverage needed to either measure or constrain the amplitude of the IGW signal to approximately the foreground confusion limit set by the aforementioned gravitational lensing signal.
Data recently released by Planck indicates that a sky region in the North contains less foreground contamination than any regions in the Southern sky~\cite{planckVI,planckII}.
Therefore our study focuses on this sky region, and observations would be made from Isi Station in Greenland.
The instrument can be configured to observe at frequencies of $\sim$~150~GHz or greater. 
Fig.~\ref{fig:spectra} Right shows the top-level experiment characteristics for the two instrument configurations considered in our study: the first tailored for observations in a single spectral band centered on 150 GHz, the second for observations in two bands centered on 210 and 267 GHz.
The plots in Fig.~\ref{fig:spectra} Left show the instrument's forecasted raw sensitivity for a one-year observation program. %
To isolate the faint B-mode signals, brighter foreground signals, such as Galactic dust emission, must be precisely measured and removed from the CMB maps.
The various sky signals may be disentangled by observing a low-dust region in the Northern Hemisphere in three spectral bands, and then fitting a frequency-dependent parametric dust model to the data pixel-by-pixel using standard likelihood techniques\cite{eriksen}.
The instrument described here is designed to be capable both of measuring the E-mode and B-mode polarization signals as well as providing high-sensitivity measurements of CMB foreground signals (see Fig.~\ref{fig:spectra}).
An overview of the instrument is described in Section~\ref{sec:instrument}, the detector system is described in Section~\ref{sec:array}, calibration and control of systematic errors are described in Section ~\ref{sec:systematics}, and the observation program is described in Section~\ref{sec:observations}.

%%%%%%%%%%%%%%%%%%%%%%%%%%%%%%%%%%%%%%%%%%%%%%%%%%%%%%%%%%%%%

\begin{figure}[t]
\centering
\includegraphics[width=\textwidth]{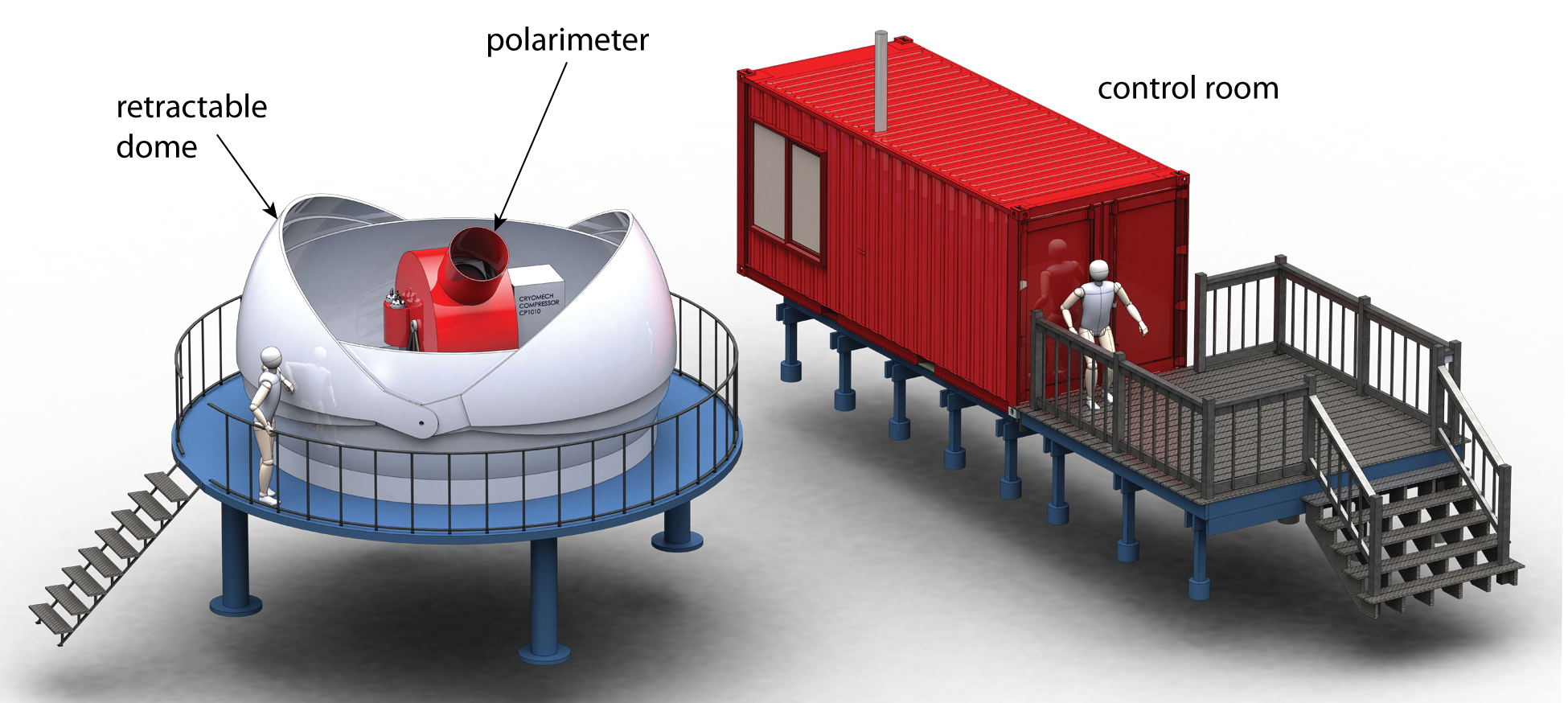}
\caption{Conceptual overview  of the telescope and observatory design.}
\label{fig:observatory}
\end{figure} 

%%%%%%%%%%%%%%%%%%%%%%%%%%%%%%%%%%%%%%%%%%%%%%%%%%%%%%%%%%%%%

%%%%%%%%%%%%%%%%%%%%%%%%%%%%%%%%%%%%%%%%%%%%%%%%%%%%%%%%%%%%%

\begin{figure}[t]
%\centering
\begin{center}
\begin{tabular}{cc}
\includegraphics[height=0.48\textheight]{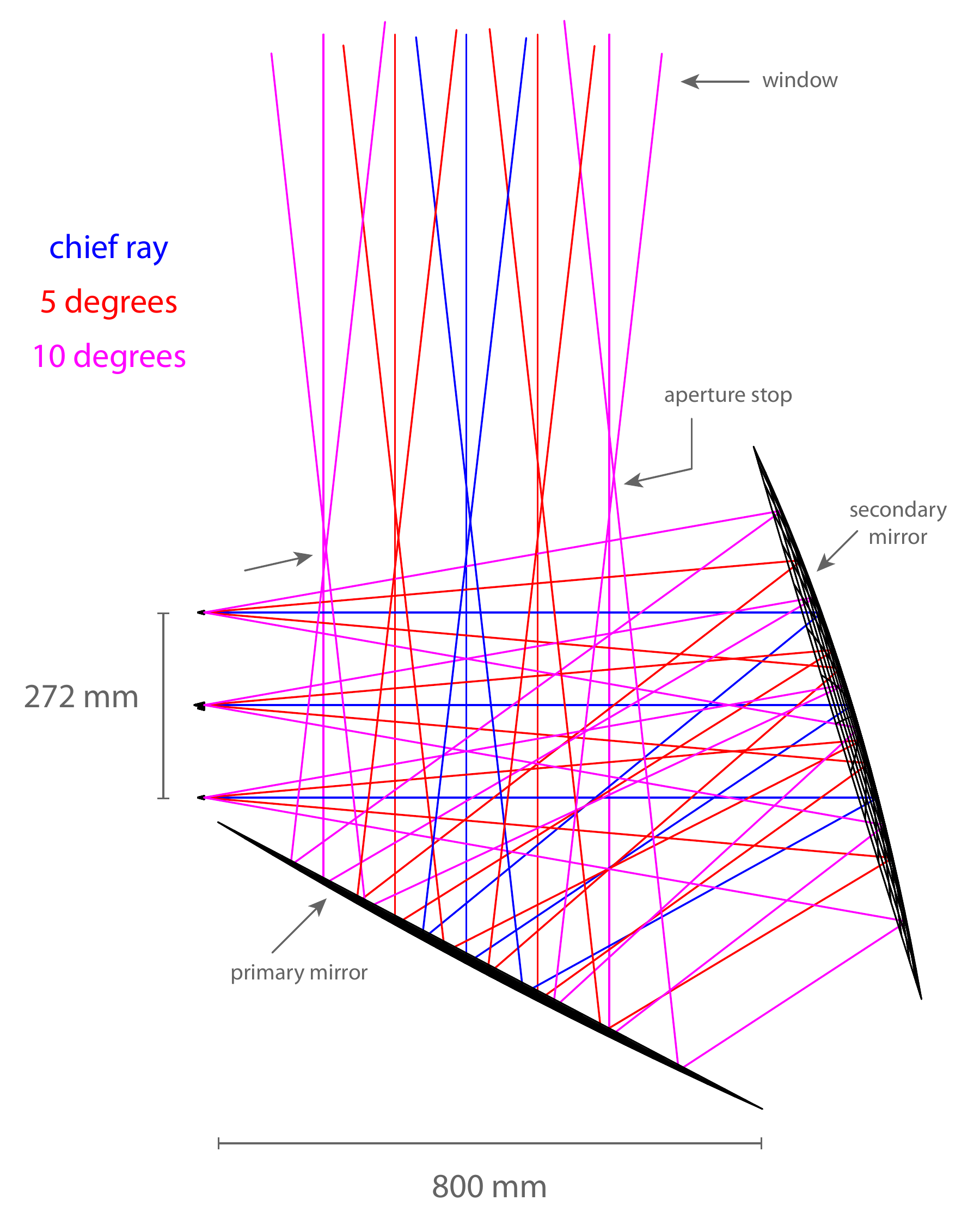} & 
\includegraphics[height=0.48\textheight]{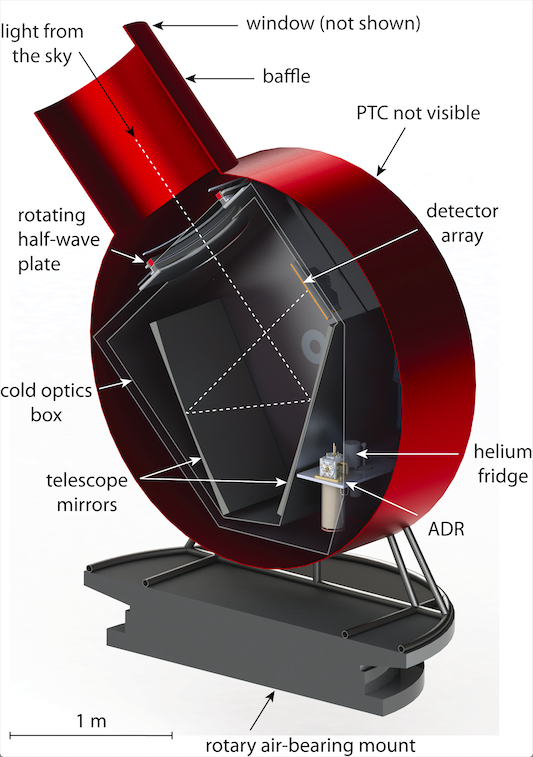}
\end{tabular}
\end{center}
\caption[mirrors]{\label{fig:mirrors} 
\textbf{Left:} Ray diagram illustrating the optical path from the window to the focal plane.
\textbf{Right:} Cross-sectional view of the receiver mounted on the precision air bearing rotary mount.  The cryostat window, the pulse tube cooler (PTC) and the helium compressors are not shown for clarity.}
\end{figure} 

%%%%%%%%%%%%%%%%%%%%%%%%%%%%%%%%%%%%%%%%%%%%%%%%%%%%%%%%%%%%%

\section{THE INSTRUMENT}
\label{sec:instrument}  

An overview of the polarimeter and the observatory is shown in Fig.~\ref{fig:observatory}. The instrument design is based on a 500 mm aperture F/2.4 catoptric Crossed-Dragone telescope, comprising an off-axis parabolic primary mirror (projected diameter 0.8 m) and a hyperbolic secondary mirror, yielding an effective focal length of 1200 mm.
The 500 mm aperture diameter produces a beam FWHM of 15~arcmin at 150 GHz .
A diagram showing the optical path from the window to the focal plane appears in Fig.~\ref{fig:mirrors} Left.
The entire optical assembly is mounted inside a cryostat (see Fig.~\ref{fig:mirrors} Right).
The instrument's optical design employs only reflecting optics held at $<$~4~K, ensuring that the dominant loading comes from the atmosphere and the window.
The telescope optics are cooled to $<$~4~K by a Chase Research Cryogenics Ltd.\ $^4$He sorption fridge backed by a Cryomech PT-415 pulse tube cooler (PTC).
The resultant total loading is approximately 3~pW at 150 GHz.

Other optical elements include a 500 mm low-pass metal-mesh filter stack that rejects high-frequency radiation from the sky; 
a rapidly rotating metal mesh half-wave plate (HWP) that modulates linearly polarized sky signals into an identifiable signal frequency band; 
and a retractable polarizer at the aperture stop to precisely calibrate cross-polarization effects.  
The first polarimeter element is the 500 mm metal-mesh HWP\cite{zhang} mounted at the telescope aperture stop.
The HWP design is sufficiently broadband to yield high performance over the targeted frequencies of each focal plane (130 - 170 GHz and 195 - 296 GHz).
The performance of a prototype metal mesh HWP is given in Zhang et al.\cite{zhang} An improved version has been developed by the Ade group with 180 degree retardance across the entire band, greater than 95\% transmission, and less than 2\% absorption.
In order to minimize vibrations and eliminate heating from stick-slip friction, the HWP is designed to rotate on a superconducting magnetic bearing (SMB) at approximately 10~Hz.
Optical systematic errors are reduced to negligible levels through rapid modulation performed on the sky side of all imaging optical elements.

The focal plane design features a detector array containing at least 2,300 lumped-element kinetic inductance detectors (LEKIDs) maintained at 100 mK with an adiabatic demagnetization refrigerator.  
The detector array is cooled to 100~mK using an adiabatic demagnetization refrigerator (ADR) and a second dedicated PTC.
Optical loading is mitigated predominantly by three design elements: a Steelcast-coated\cite{fixsen} optics box, held at $<$~4~K; a radiation shield in the cryostat snout, held at $\sim$50~K, designed to manage far side-lobes; and a 500~mm diameter low-pass metal-mesh filter stack that rejects high-frequency radiation from the sky (see Fig.~\ref{fig:mirrors} Right).

The instrument utilizes a commercially available heavy-duty rotary table floating on a precision air bearing, specifically designed to work in the low-temperature and low-pressure environment in Greenland.
The mount is capable of rotating 360$^\circ$ in azimuth.
The mount design also includes inclinometers and a mechanical leveling system to ensure that the table remains level in case of drifting or settling on the Greenland observation site's ice bed.
Slip rings provide AC power, a wired Ethernet connection for instrument control and data retrieval, and cooling fluid for the helium compressors.
The helium compressors are mounted on the rotating platform.
The mount includes an inductive position encoder, which is used by the rotary servo loop.
The instrument is designed to rotate in azimuth at a series of fixed elevation angles, tracing circles on the sky.
The mount is capable of rotating faster than 5$^\circ$/s in azimuth, which is the requirement to meet the desired scan speed of 2$^\circ$/s on the sky.
The entire instrument is compact, designed to be fully tested in the laboratory and shipped in one piece to the observation site, thereby eliminating the need to perform a complicated and error-prone on-site assembly procedure.   
Required pointing accuracy is set at 12" by the requirement that the induced B-mode error from pointing reconstruction will be significantly less than 10\% of the anticipated lensing and gravitational wave B-mode for all relevant $\ell$ values, to a value\footnote{The tensor-to-scalar ratio $r$ that quantifies the amplitude of the primordial B-mode signal is defined as $r$ $\equiv r_{0.002} = A_t/A_s$, where $A_t$ is the primordial spectrum of tensor fluctuations associated with the primordial B-mode signal at a scale of $k$ = 0.002 Mpc$^{-1}$, and $A_s$ is the corresponding spectrum of scalar fluctuations associated with the signal produced by density inhomogeneities.  A critical value of $r$ is $r \sim 0.01$, at which point the energy scale of inflation reaches that relevant for Grand Unified Theories (GUTs).  See, e.g., Baumann et al.\cite{baumann2009} for a recent review.} of $r=0.01$ \cite{hhz03}.
Absolute pointing may be constructed to at least the required accuracy via information from optical encoders calibrated with pointing models built from daily measurements taken with a custom-built star camera\footnote{The star cameras built by the Miller group for the EBEX experiment have a pointing accuracy of ~1". See Reichborn-Kjennerud et al.~\cite{ebex}}.
%

%%%%%%%%%%%%%%%%%%%%%%%%%%%%%%%%%%%%%%%%%%%%%%%%%%%%%%%%%%%%%

\begin{figure}[t]
\begin{center}
\begin{tabular}{ccc}
\includegraphics[height=0.3\textheight]{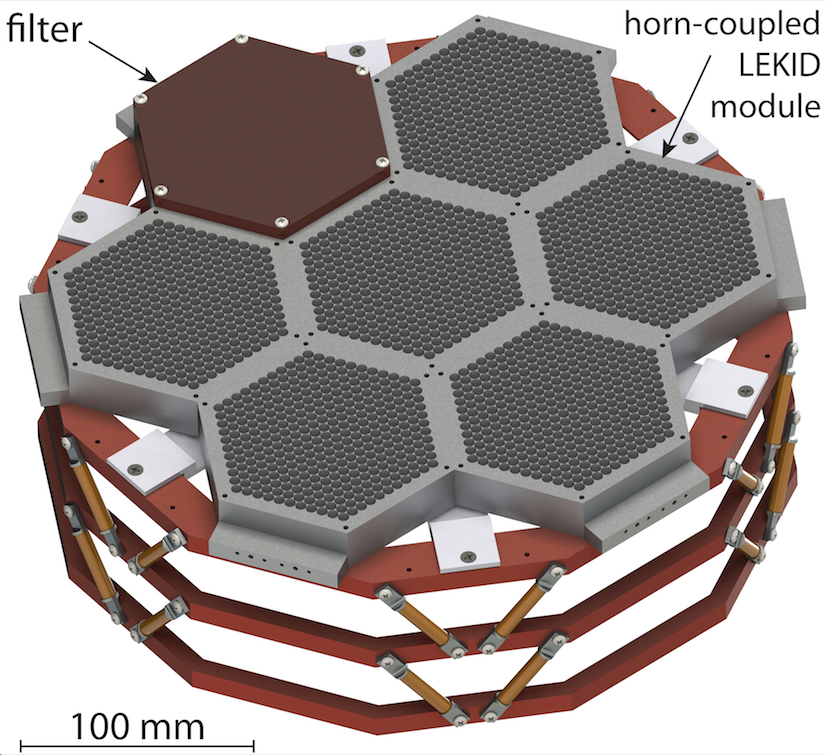} & &
\includegraphics[height=0.3\textheight]{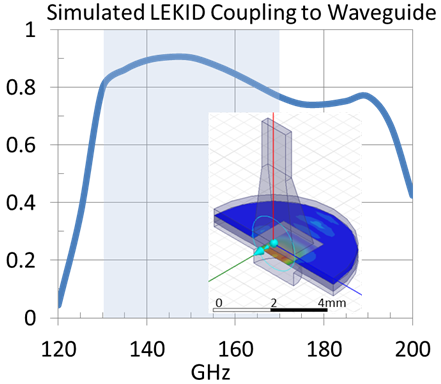}
\end{tabular}
\begin{tabular}{ccc}
\includegraphics[height=0.3\textheight]{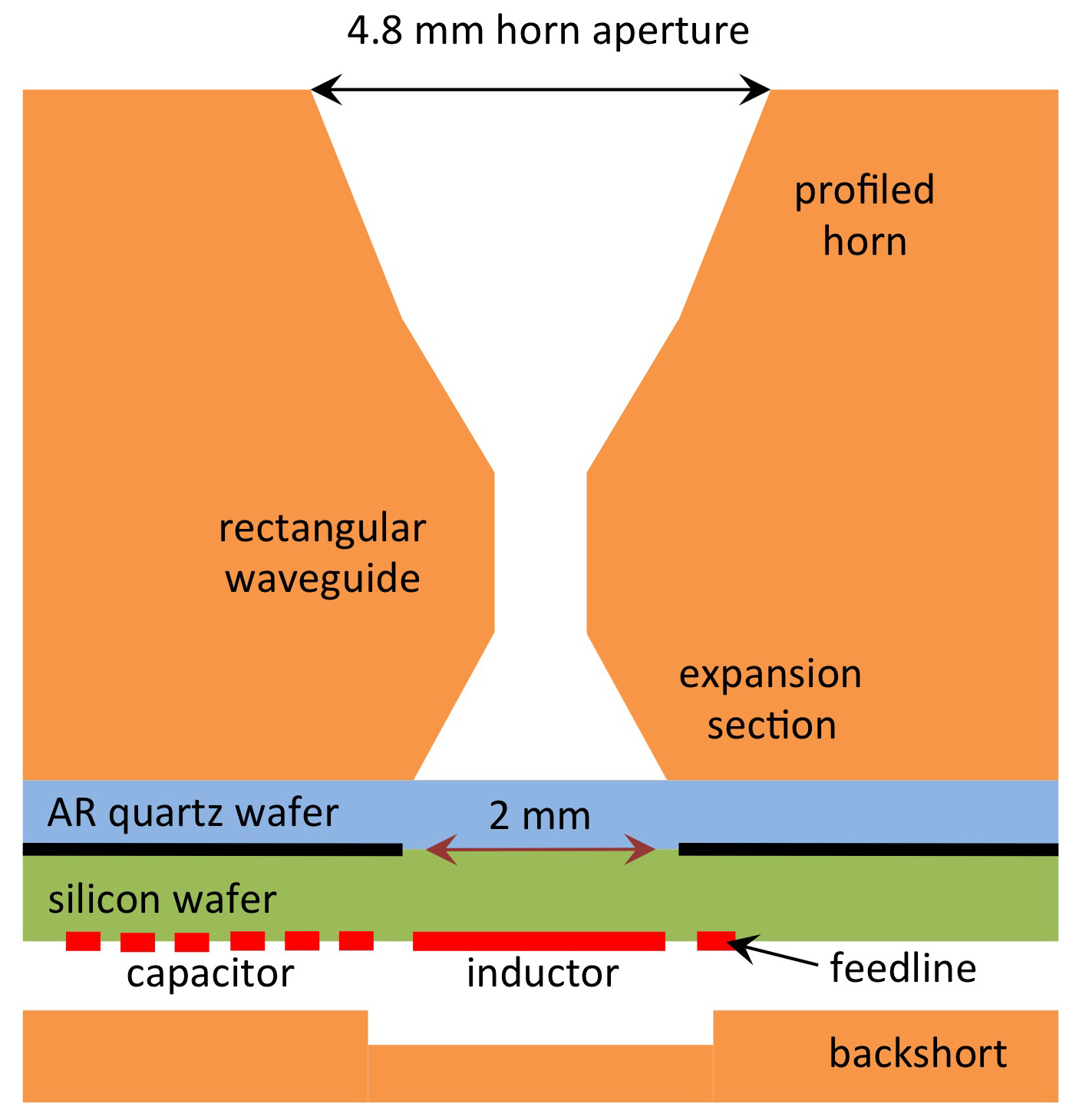} & &
\includegraphics[height=0.3\textheight]{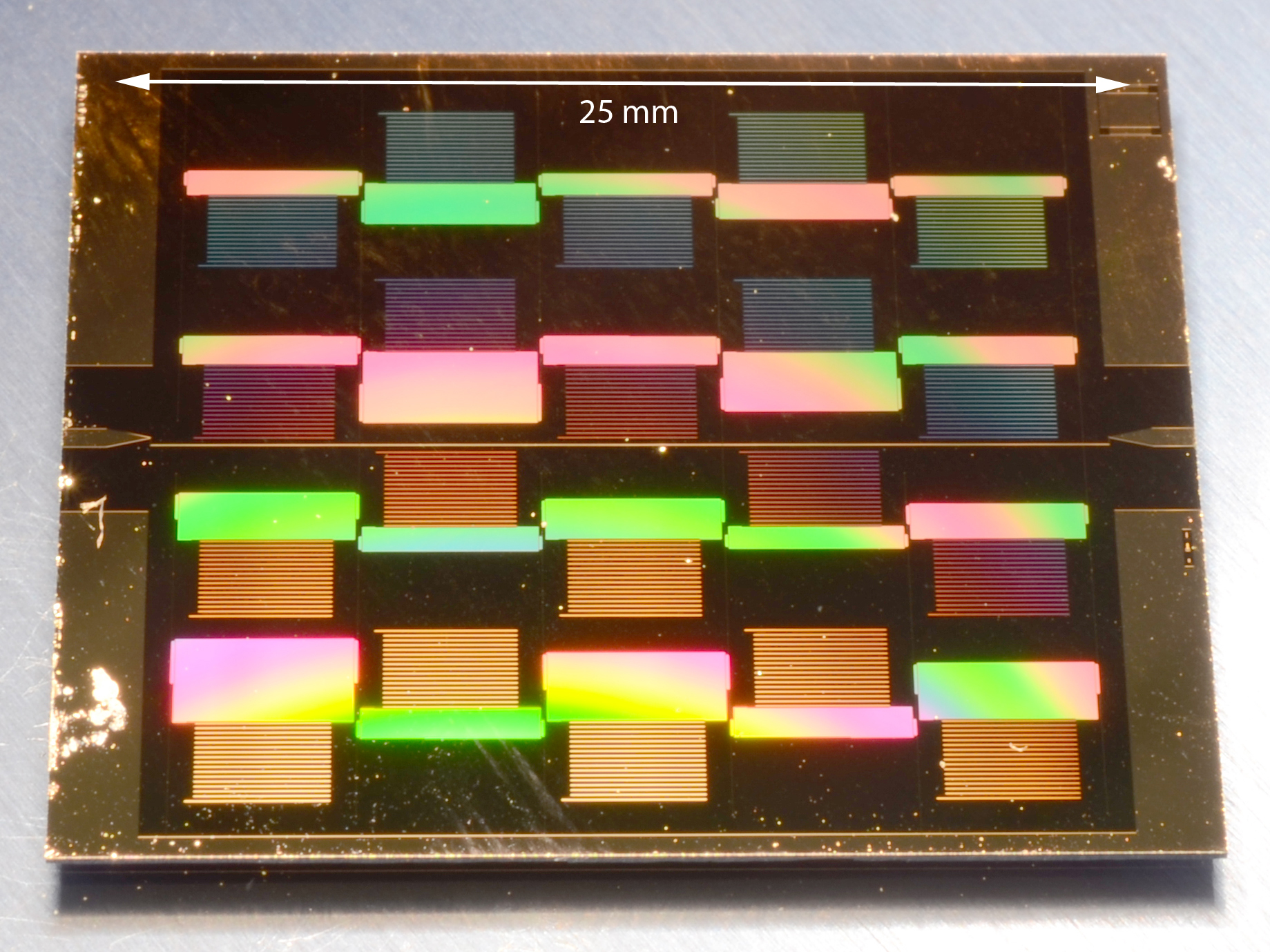}
\end{tabular}
\end{center}
\caption[kids] 
%>>>> use \label inside caption to get Fig. number with \ref{}
{ \label{fig:kids} 
\textbf{Top Left:} The detector array at the focal plane consists of seven hexagonal modules.  For the 150~GHz spectral band, each $\sim$100 mm diameter module contains 331 horns yielding a total of 2,317 LEKIDs in the single-polarization array. \textbf{Top Right:} Results from an HFSS electromagnetic simulation (inset) showing the power coupling efficiency for the horn-coupled LEKID array with an AR coating. The 150~GHz band is shown in gray. \textbf{Bottom Left:} Schematic cross-section through one of the pixels (not to scale).  The 20 nm thick Aluminum LEKID is shown in red and patterned 20 nm thick TiN is shown in black.
\textbf{Bottom Right:} A photograph of a 20-element prototype LEKID wafer optimized for CMB observations that was designed by our collaboration and fabricated at JPL.
}
\end{figure} 
  
%%%%%%%%%%%%%%%%%%%%%%%%%%%%%%%%%%%%%%%%%%%%%%%%%%%%%%%%%%%%%

\begin{figure}[t]
\begin{center}
\begin{tabular}{cc}
\includegraphics[height=0.3\textheight]{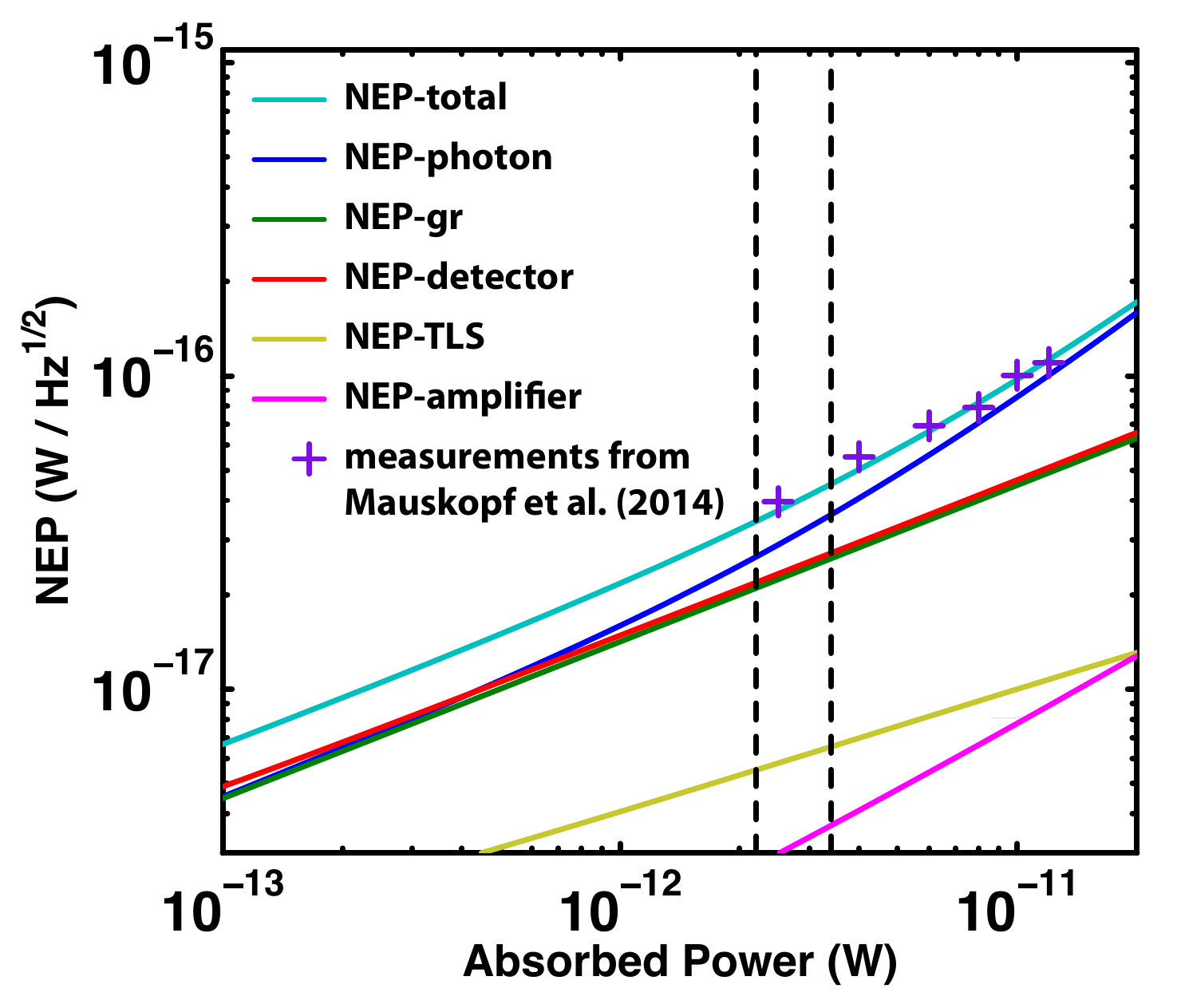} &
\includegraphics[height=0.28\textheight]{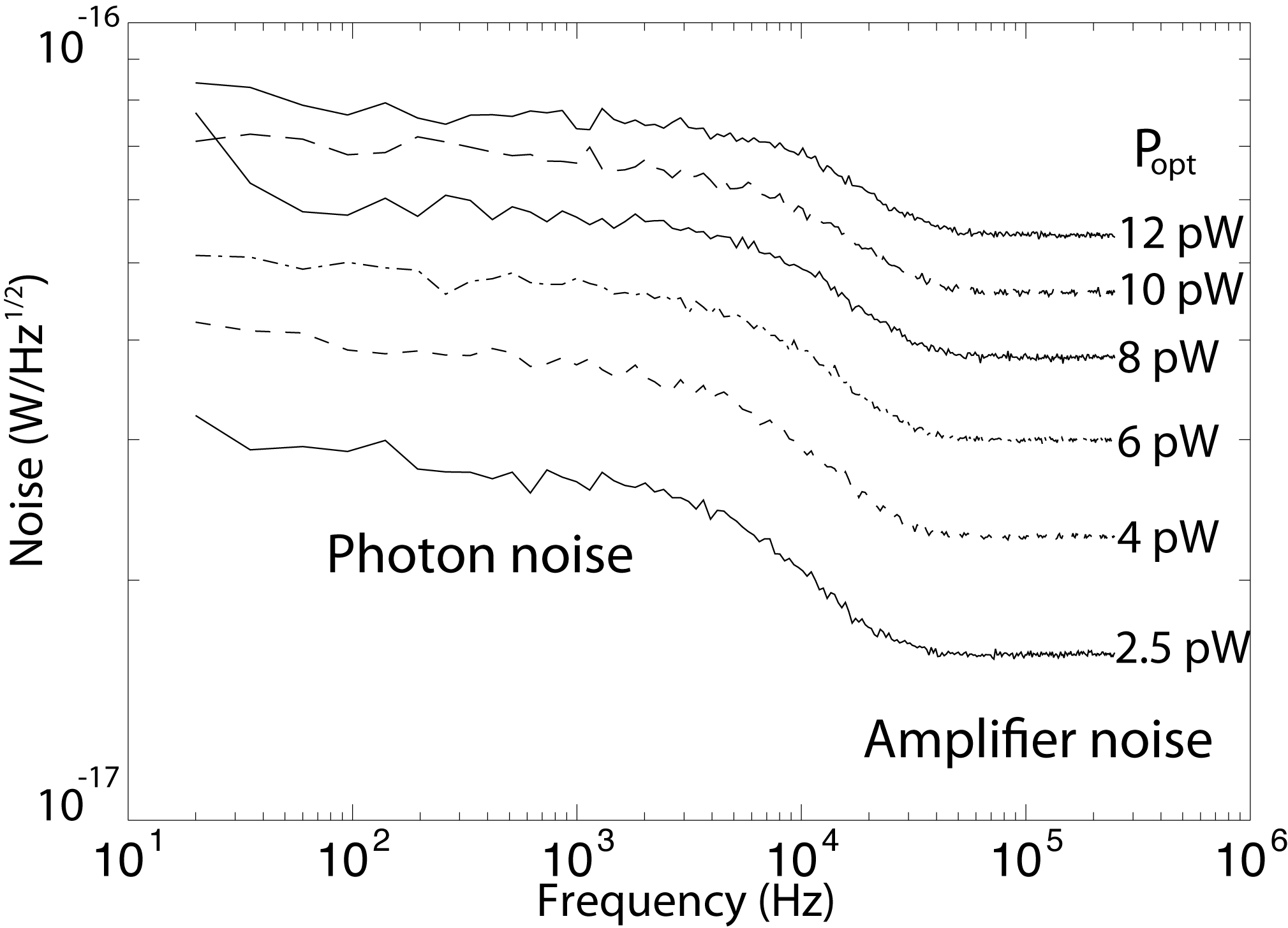}
\end{tabular}
\end{center}
\caption[nep] 
%>>>> use \label inside caption to get Fig. number with \ref{}
{ \label{fig:nep} 
\textbf{Left:} Projected LEKID performance. The ``detector NEP'' curve includes contributions from g-r noise, TLS noise, and amplifier noise, while the ``total NEP'' curve includes the added contribution from photon noise. The expected optical loading at zenith angles between 15$^\circ$ and 60$^\circ$ in Greenland is indicated by the vertical dashed lines.  The cross symbols are measured points from the data displayed in the Right panel.  Because $Q_r \sim 2.2\times10^{4}$, all 331 detectors on one module may be multiplexed in a 140-210~MHz band on a single RF transmission line. \textbf{Right:} Measured noise spectra from a single LEKID under different loading conditions.  These measurements were made using a cryogenic test bed at Cardiff with an internal blackbody source \cite{mauskopf}. The noise below 1~kHz agrees with the expected combination of g-r noise and photon noise and rolls off around 10-20~kHz due to the quasiparticle lifetime. The noise level above 50~kHz corresponds to amplifier noise and increases in NEP units with optical loading due to both the reduction in the quality factor of the resonator and the decrease in the quasiparticle lifetime. 
}
\end{figure} 

%%%%%%%%%%%%%%%%%%%%%%%%%%%%%%%%%%%%%%%%%%%%%%%%%%%%%%%%%%%%%

\section{THE DETECTOR ARRAY} 
\label{sec:array}

Kinetic inductance detectors (KIDs) consist of superconducting resonators that change their resonant frequencies and quality factors in response to incident Cooper pair breaking photons; see Zmuidzinas\cite{zmuidzinas} for an in-depth treatment.
We chose to use KIDs for this study because they are inherently multiplexable, they are relatively simple to fabricate and integrate in an instrument, and their noise properties are maturing.
These advantages have spurred a rapid growth of activity.
At present, $\sim 20$~groups are actively developing KIDs worldwide.

%%%%%%%%%%%%%%%%%%%%%%%%%%%%%%%%%%%%%%%%%%%%%%%%%%%%%%%%%%%%%

\subsection{Aluminum LEKIDs}

The focal plane design consists of an array of LEKIDs that are suitable for use with feed horns and designed with conservative geometries that require only 2-3 lithography steps for fabrication.
The left panel of Fig.~\ref{fig:nep} shows our theoretical performance estimates for LEKIDs fabricated from 20~nm thick aluminum films deposited on silicon substrates.
At the design-specified operating point of $P_{absorbed} \approx 3$~pW, the detector NEP should be dominated by recombination noise.

Two-level system (TLS) noise has been a challenge for KID designs.
TLS noise is caused by a surface layer of dielectric fluctuators, in which the associated capacitance fluctuations contribute to detector noise by perturbing the resonator frequency~\cite{zmuidzinas,kumar,gao,gao2,gao3,gao4,noroozian}.
Although TLS noise dominated many early KID designs, an improved understanding of the underlying physics has enabled the engineering of LEKID capacitors that reduce the TLS noise contribution below the fundamental photon noise limit~\cite{gao3,noroozian}.
Reports of photon noise limited operation have been published with sensitivities in the range NEP $\sim$ 1 to 50 $\times$ 10$^{-17}$~W/$\sqrt{\mbox{Hz}}$~\cite{yates,mckenney}.
The right panel of Fig.~\ref{fig:nep} shows the measured noise spectrum of a photon noise limited LEKID designed and fabricated by members of our collaboration.

%%%%%%%%%%%%%%%%%%%%%%%%%%%%%%%%%%%%%%%%%%%%%%%%%%%%%%%%%%%%%

\begin{figure}[t]
\begin{center}
\begin{tabular}{cc}
\includegraphics[height=0.25\textheight]{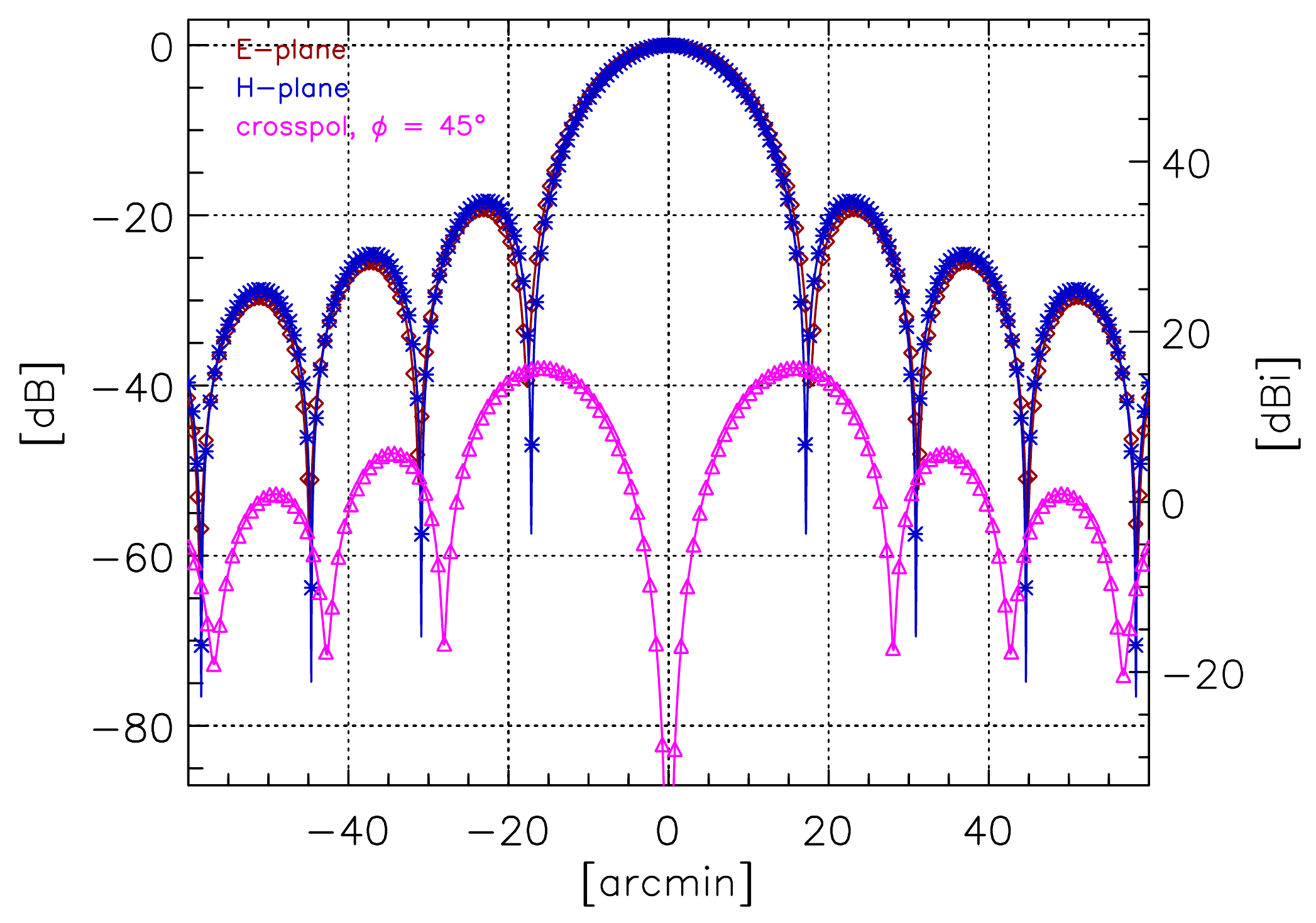} &
\includegraphics[height=0.25\textheight]{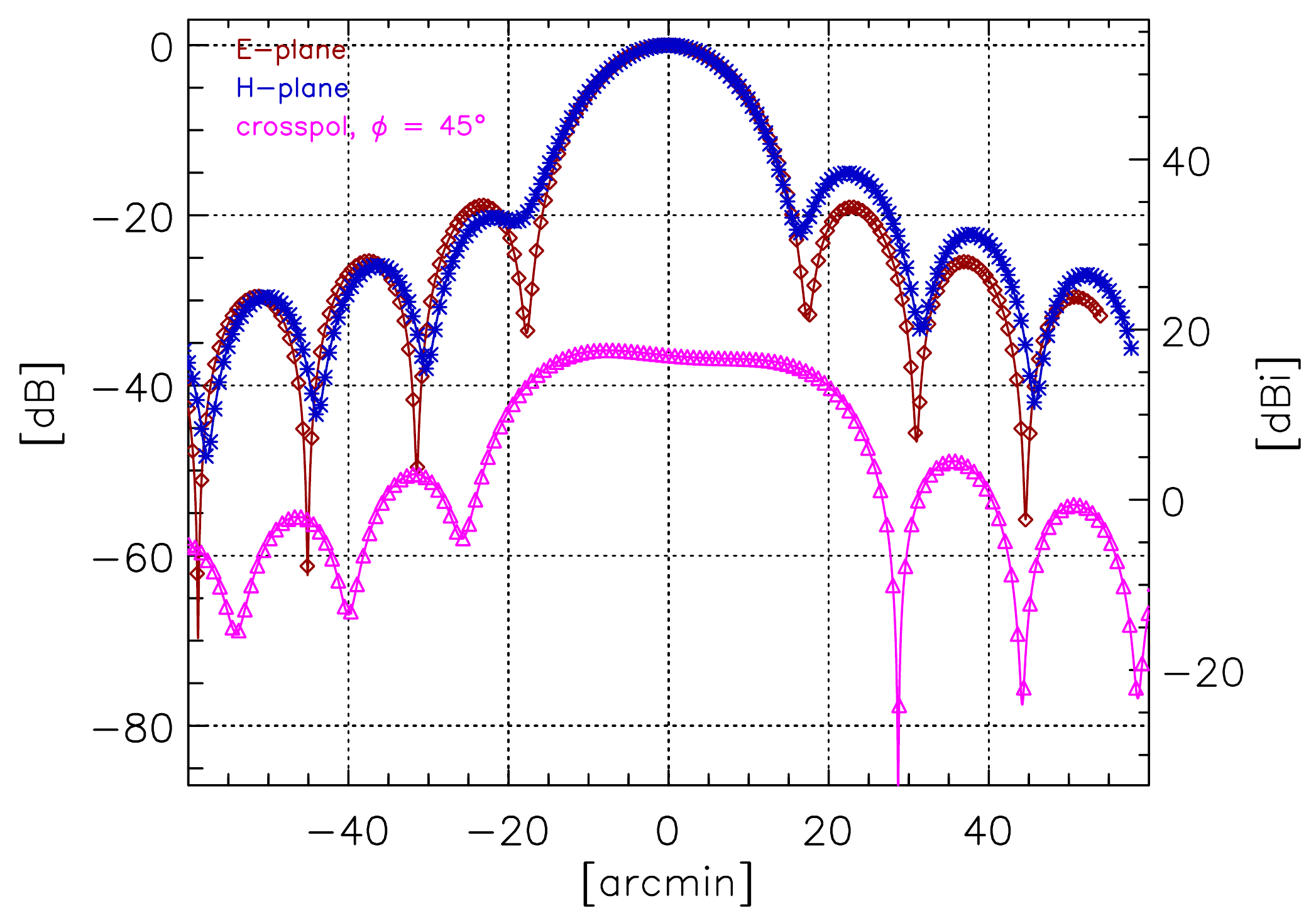} 
\end{tabular}
\end{center}
\caption{\textbf{Left:} Simulated center feed beam cuts for 150~GHz. \textbf{Right:} Simulated edge beam cuts for 150~GHz.  These simulations use a simple conical horn.  They show the best and worst case performance for this optical system.}
\label{fig:far_side_lobe}
\end{figure}

%%%%%%%%%%%%%%%%%%%%%%%%%%%%%%%%%%%%%%%%%%%%%%%%%%%%%%%%%%%%%

The measured noise spectrum of a LEKID designed and fabricated by members of our collaboration, demonstrating photon noise limited performance, is shown in the right panel of Fig.~\ref{fig:nep}.  It is in excellent agreement with predictions (see Fig.~\ref{fig:nep} Left). A photo of a prototype array designed by our collaboration and fabricated at the Jet Propulsion Laboratory is shown in
Fig.~\ref{fig:kids} Bottom Right.

Each instrument configuration includes a $\sim$300 mm diameter detector focal plane, containing seven hexagonal modules (see Fig.~\ref{fig:kids} Top Left).  A single $\sim$100 mm diameter module contains 331 horn-coupled LEKIDs, for a total of 2,317 detectors. All detectors within a single module are frequency-multiplexed into a 140-210 MHz readout band, thus making the entire seven-module focal plane readable using only seven SiGe bipolar cryogenic low noise amplifiers (LNAs)\cite{bardin} and seven pairs of coaxial cables. We have developed the detector readout electronics from the successful Open Source Readout system developed at Caltech for the MUSIC instrument\cite{golwala}. The design is based around the ROACH signal processing board developed by the CASPER collaboration\footnote{For further information see the CASPER collaboration website: http://casper.berkeley.edu.}, which hosts a Xilinx Virtex 5 field-programmable gate array (FPGA). Multiplexing at even higher density and at the full system level have been demonstrated as part of the MAKO project\cite{swenson}.

%%%%%%%%%%%%%%%%%%%%%%%%%%%%%%%%%%%%%%%%%%%%%%%%%%%%%%%%%%%%%

\subsection{Optical coupling}

The optical coupling design includes profiled horns to obtain optimized circular beams on the sky.  The use of horns to couple the detectors to the telescope provides a number of advantages over bare-pixel LEKID arrays.  Space is made available for a large interdigital capacitor, thereby lowering the readout frequency and reducing the effects of TLS noise; spillover and stray light inside the optics box are reduced; and an integrated high pass filter and a high performance polarization selection are provided by a section of rectangular waveguide. The relatively small horn size of 1F$\lambda$ = 4.8 mm is chosen to maximize mapping speed\cite{griffin}, thus exploiting the high multiplexing density and high sensitivity associated with LEKIDs. The horns couple with approximately 39\% efficiency to the sky, while the remaining 61\% of the horn beam is terminated on the $<$~4~K shield.

Simulation and optimization techniques have recently been developed to design smooth-walled horns that are easily drillable using custom shaped tools \cite{Yassinetal2007,Zeng2010,Leech2012,Tan2012} and provide circular beams on the sky with low cross-polarization. The profile of a single horn consists of several flat or conical sections that provide an optimized field pattern at the mouth of the horn similar to the Pickett-Potter horn, \cite{Potteretal1961} but with wider bandwidth and an easier manufacturing process than continuously-curved taper designs \cite{Nielsonetal2002}. The THz laboratory at ASU has a high-precision micromilling machine (Kern 44 with 0.5 $\mu$m position accuracy) that has been used to fabricate several versions of these types of horns for use in ground-based high frequency heterodyne instruments such as SuperCAM \cite{Groppi2010}. The instrument horns will be easier to machine because they operate at significantly longer wavelengths. 

As detailed in Fig.~\ref{fig:kids} Bottom Left, the horn narrows to a single-mode rectangular waveguide section to achieve polarization selection and definition of the low frequency band edge. The waveguide then re-expands to flatten the propagation impedance at the low frequency edge of the band, improving optical coupling and allowing the radiation to launch efficiently into the silicon LEKID wafer through a quartz anti-reflection (AR) wafer. The baseline pixel design consists of a back illuminated single-polarization LEKID in which the inductor/absorber is a meandered aluminum trace on a silicon substrate with a filling factor designed to match the wave impedance.  
This design is similar to the absorbers used in the NIKA\cite{monfardini2} and MAKO\cite{swenson} instruments. A quarter-wavelength antireflection (AR) layer of quartz is used for impedance matching to the silicon wafer. The high refractive index of silicon reduces the wavelength by a significant factor, $\lambda / \sqrt{\epsilon_r} \approx 0.29 \lambda$.  HFSS simulations show that the radiation launched from the waveguide remains well collimated as it travels through the 300 $\mu$m thick wafer.  These simulations predict that an inductor/absorber of area $\lambda$ $\times$ $\lambda$ $\approx$ 4 mm$^2$ (see Fig.~\ref{fig:kids} Top Right) has an absorption efficiency of $\sim$ 90\% over the 125 to 175~GHz spectral band. 

The AR and detector wafers are mounted to the back of the horn plate and a metal back plate is attached to seal each module. The back plate includes metal cavities behind each detector designed to act as backshorts $\sim$~$\lambda$/4 in length. To prevent scattered radiation from coupling to adjacent detectors, the side of the silicon wafer facing the horns is metallized with titanium nitride (TiN) between detectors.  The design includes holes in the TiN layer to allow radiation to propagate from the horns through the quartz and silicon to the detectors (see Fig.~\ref{fig:kids} Bottom Left).  This metallization acts simultaneously as an efficient mm-wave absorber with an effective sheet resistance of approximately 150 $\Omega$, and as an absorber of ballistic phonons propagating in the silicon.  Optical crosstalk between pixels ($<$ 1\%) is projected to be comparable to other well-designed detector systems used for CMB polarimetry, and may be accounted for in post-observation analysis by use of an appropriate pointing matrix.

The baseline design for the optical coupling described above is simple and requires no new detector development.  Although this design is capable of detecting only a single polarization per pixel, we have also designed a more complex dual-polarization pixel similar to the dual-polarization 90 GHz TES modules used in SPTPol\cite{crites} and the PoleKID detectors being developed for the SuperBLASTPol balloon-borne sub-mm telescope\cite{hubmayr}.  This would double the number of detectors in the focal plane.

%%%%%%%%%%%%%%%%%%%%%%%%%%%%%%%%%%%%%%%%%%%%%%%%%%%%%%%%%%%%%

\begin{figure}[t]
\begin{center}
\begin{tabular}{c}
\includegraphics[height=0.35\textheight]{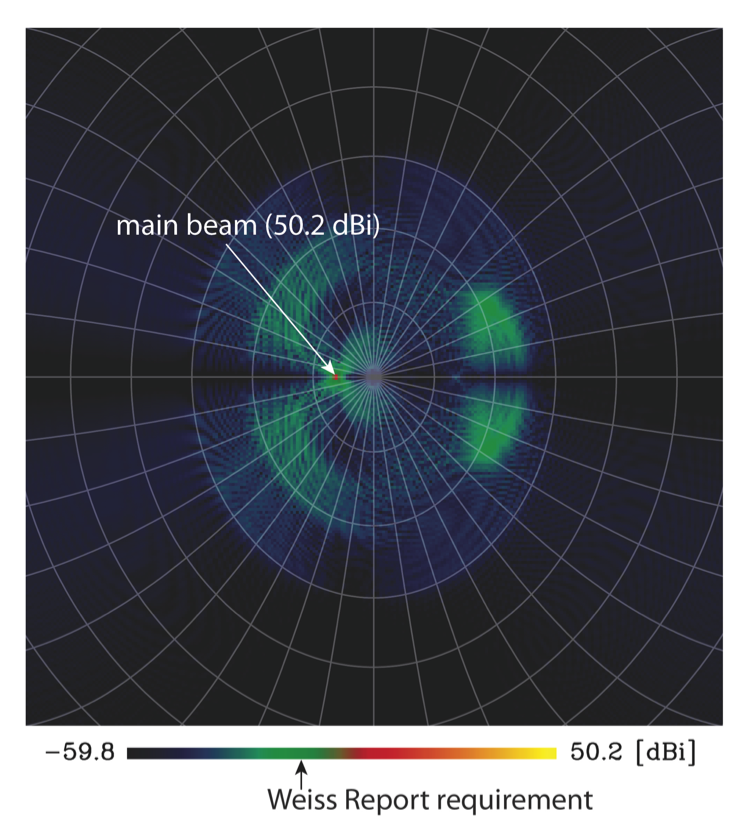}
\end{tabular}
\end{center}
\caption{An electromagnetic simulation of the far side lobe response for a single pixel at the edge of the focal plane. Graticule lines are spaced by $10^{\circ}$. Far side lobes are suppressed to less than -100~dB beyond a radius of 30$^{\circ}$ due to the $< 50$~K radiation shield in the cryostat snout. Within a radius of 30$^{\circ}$, the far side lobe levels are less than -70~dB.}
\label{fig:far_side_lobe2}
\end{figure}

%%%%%%%%%%%%%%%%%%%%%%%%%%%%%%%%%%%%%%%%%%%%%%%%%%%%%%%%%%%%%

\section{Calibration and Control of Systematic Errors} 
\label{sec:systematics}

Systematic errors are reduced to negligible levels by the combination of a 15 arcminute beam, a cryogenically cooled, high performance optical system, a rapidly rotating cold HWP as the first polarimeter element, and a retractable polarizer at the aperture stop designed to minimize the effects of cross-polarization to better than 0.1$^\circ$.
The optical system design includes four features that minimize systematic errors: (i) emission and side lobe levels are reduced dramatically by fully containing the optics within the $< 4$~K optics box; (ii) the cold reflective re-imaging optics produce low levels of instrumental polarization and cross-polarization; (iii) the horns and aperture stop create high quality detector beams across the entire focal plane; and (iv) the continuously rotating half-wave plate averages down time-independent effects and mitigates low-frequency drifts.
Table~\ref{tab:systematics} summarizes conservative requirements for systematic errors from the CMB literature, systematics expected from the optical system alone\footnote{The optical system systematics were modeled using the GRASP, a physical optics software package: http://www.ticra.com/products/software/grasp.} (including the mirrors, cold aperture stop, and feed horns), and the levels below which they will be negligible due to rotation of the HWP and/or calibration. The given limit for each systematic error would produce approximately 10\% of the B-mode signal corresponding to $r = 0.01$.

The continuously rotating HWP plays a major role in mitigating systematic errors. A HWP rotating at frequency $f$ modulates polarized signals into sidebands of $4 f$, allowing signals outside the sky signal band $4 f \pm \Delta$ to be rejected in analysis.  The design specified scan speed is roughly $2^\circ$/s on the sky.  When combined with the $0.25^\circ$ detector beam, this scan speed creates a low pass filter and sets the demodulation bandwidth at about $\Delta = 8 \, \mathrm{Hz}$.  The design includes a HWP rotating at approximately 10~Hz.  This causes the polarized sky signal to appear in  time-ordered data between 32 and 48~Hz, where the instrumental noise is expected to be white (see Fig.~\ref{fig:nep}).
Maps of the detector beams with and without the polarizer may be used to model and account for any non-idealities in the HWP. 
Simulations show that rapid modulation performed on the sky side of all imaging optical elements reduces optical systematic errors to negligible levels (see~Table~\ref{tab:systematics}). 
For example, differential HWP transmittance \cite{brown} and polarized emission produce signals at $2 f = 20 \, \mathrm{Hz}$, outside the signal band. 

Because all optical components are on the detector side of the HWP, many traditional sources of instrumental systematic errors are reduced to negligible levels, as signals appearing at frequencies below the modulation frequency appear outside the signal bandpass.   
Examples of sources of error that are mitigated this way include: scan-synchronous temperature signals capable of generating polarized emission from optical elements; cryogenic temperature fluctuations; differential reflection of orthogonal polarizations from mirrors; and cross-polar response in the detectors. Quantitative estimates of the expected error mitigation for these effects appear in Table~\ref{tab:systematics}.
Reflection of stray light is minimized by coating the cold optics box with the absorbent material Steelcast \cite{fixsen}.  Our simulation of the far side lobe response (see Figs.~\ref{fig:far_side_lobe} and \ref{fig:far_side_lobe2}) shows that its performance exceeds the specification.  Instrumental polarization and cross-polar response generated on the sky-side of the HWP are modulated in the same way as the sky signal. The dielectric window and IR-blocking filters, the only optical elements on the sky side of the HWP, are negligible contributors of polarized systematic effects.

%%%%%%%%%%%%%%%%%%%%%%%%%%%%%%%%%%%%%%%%%%%%%%%%%%%%%%%%%%%%%

\begin{table}[t]
\caption{Contamination from CMB temperature anisotropies is dT $\rightarrow$ B; contamination from E-modes is E $\rightarrow$ B. Pointing error refers to the limit on reconstructed pointing, not real-time pointing. For limits on systematic errors marked $\star$ see the Weiss Report \cite{weiss}; for $\star\star$ see Hu et al.\cite{hhz03}; for $\star\star\star$ see O'Dea et al.\cite{odea}, which defines average and differential ellipticity.}
\label{tab:systematics}
\begin{center}
\begin{tabular}{lccccc}
 & Effect  & Requirement for & Performance of  & Limit After &         \\
  & on      & Negligible Effect & Optical System  & HWP Rotation   & See     \\
Source of Systematic Error & B-Modes & to $r<0.01$       & Alone           & \& Calibration & Section \\
\hline
Instrumental Polarization$^\star$
  & dT $\rightarrow$ B
  & $10^{-4}$
  & $\sim 10^{-3}$
  & $\ll 10^{-4}$
  & \ref{sec:systematics} \\
Cross-Polarization (Q-U mixing)$^\star$
  & E $\rightarrow$ B
  & $6 \times 10^{-3}$
  &  $< 10^{-2}$
  & $< 3 \times 10^{-3}$
  & \ref{sec:systematics} \\
Polarization Angle Uncertainty$^\star$
  & E $\rightarrow$ B
  & $0.2^\circ$
  &  
  & $< 0.1^\circ$
  & \ref{sec:systematics} \\
Far Side Lobe Response$^\star$
  & dT $\rightarrow$ B
  & $-60$ dB
  & $< -70$ dB
  & $< -70$ dB
  & \ref{sec:systematics} \\
Pointing Error$^{\star\star}$ % Keep star if we use limit from HHZ.
  & E $\rightarrow$ B
  & 12"
  & 
  & $<$ 10"
  & \ref{sec:instrument} \\
Beam Average Ellipticity$^{\star\star\star}$
  & E $\rightarrow$ B
  & $< 0.15$
  & $< 0.01$
  & $< 0.01$
  & \ref{sec:systematics} \\
Beam Differential Ellipticity$^{\star\star\star}$
  & dT $\rightarrow$ B
  & $< 0.006$
  & $< 0.007$
  & negligible
  & \ref{sec:systematics} \\
\hline
\end{tabular}
\end{center}
\end{table}

%%%%%%%%%%%%%%%%%%%%%%%%%%%%%%%%%%%%%%%%%%%%%%%%%%%%%%%%%%%%%

The design employs the following calibration strategies to ensure that unwanted B-mode signals produced by instrumental and scan effects are more than an order of magnitude smaller than the primordial signal to $r<0.01$:
Direct comparison with WMAP and Planck temperature maps enables absolute calibration at the 1\% level.  Drift in the responsivity of the detector array can be monitored by periodically illuminating the detectors with a stable millimeter-wave source built into the cryostat.
The total power response of the instrument's beam may be mapped by planetary observations: by performing standard azimuth rotations near the instrument's lower elevation limit as the planet rises and sets.
The azimuth scan data would be obtained regularly during science observations. 
Fixed azimuth elevation nods may be performed periodically throughout the observational program to provide a cross-check and information on any elevation-induced systematics. Each scan through the planet is capable of providing knowledge of the beam response of each detector to the 0.01\% level, as well as the offset between the microwave beams and the fiducial boresight pointing defined by the star cameras to $<$~10~arcsec.
In addition to observing typical polarization calibrators such as Tau A during the course of the survey, the design includes a retractable wire-grid polarizer which can be inserted into the optical path inside the cryostat during calibration scans to effectively transform planetary sources and the CMB into completely linearly polarized sources. This allows scans on planetary sources to also provide polarized beam maps, as well as exquisite knowledge of the absolute polarization angle of the instrument relative to the orientation of the wire-grid polarizer, which in turn may be mechanically referenced to the boresight fiducial axis, and hence the sky.  Repeating such scans on a weekly basis would highlight any unexpected changes in the performance of the instrument.
Measuring the absolute polarization angle to better than 0.2 degrees is critical. To achieve this, the cryostat design is tailored to provide precise, repeatable alignment to better than 0.1 degrees between the HWP angular encoder fiducial angle, the cold retractable polarizer, and the boresight fiducial axis. A layer of microwave absorber is placed in front of the retractable wire-grid polarizer in the optical path before the rotating HWP and cooled to $\sim 4$~K inside the cryostat. By measuring the angle of the HWP corresponding to maximum response for each detector, one can derive both the polarization angle as a function of the HWP angle, and the polarization modulation efficiency for each detector.
%

%%%%%%%%%%%%%%%%%%%%%%%%%%%%%%%%%%%%%%%%%%%%%%%%%%%%%%%%%%%%%

\begin{figure}[t]
\begin{center}
   \begin{tabular}{cc}
   \includegraphics[height=4.4cm]{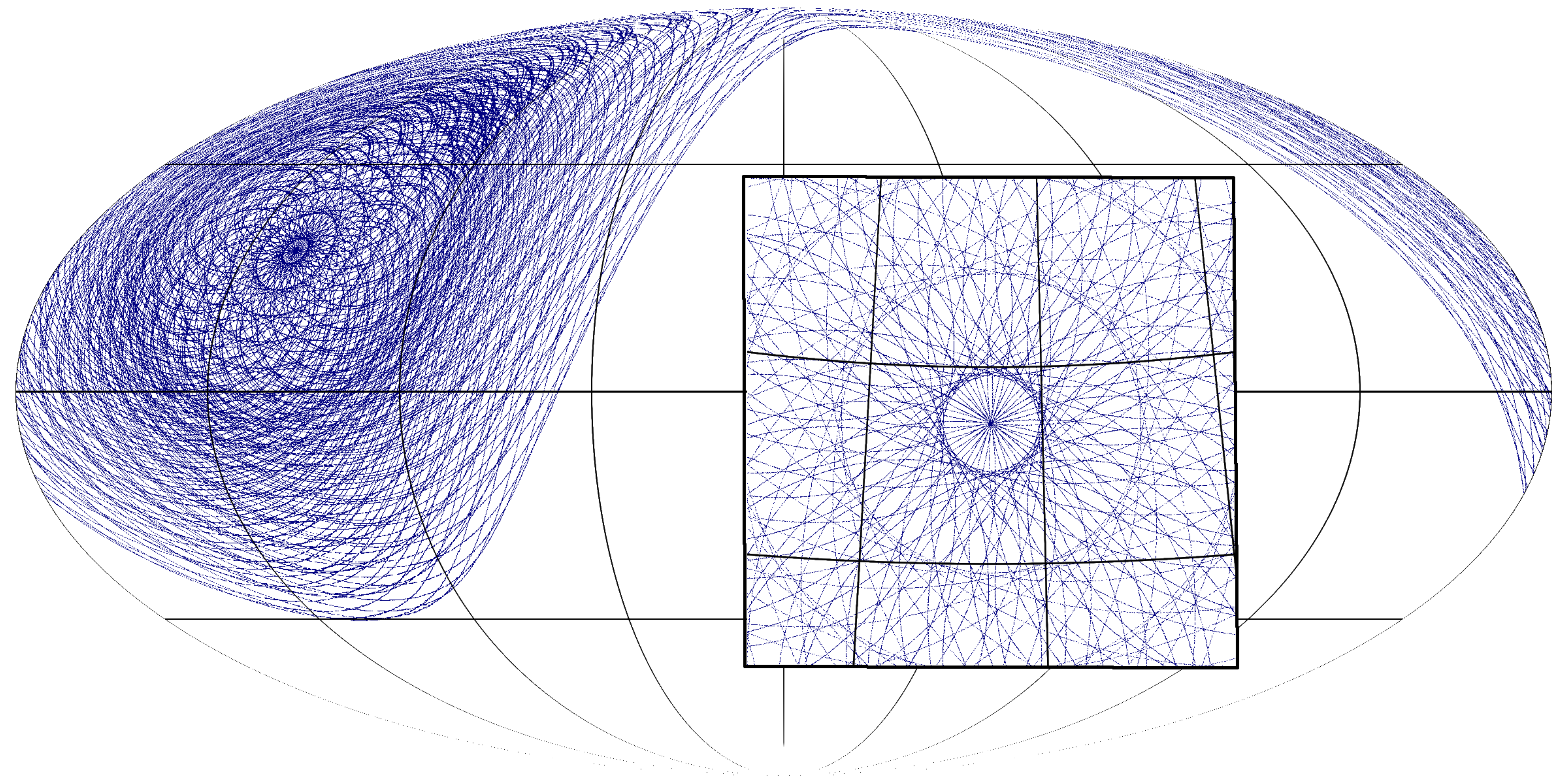} &
   \includegraphics[height=4.5cm]{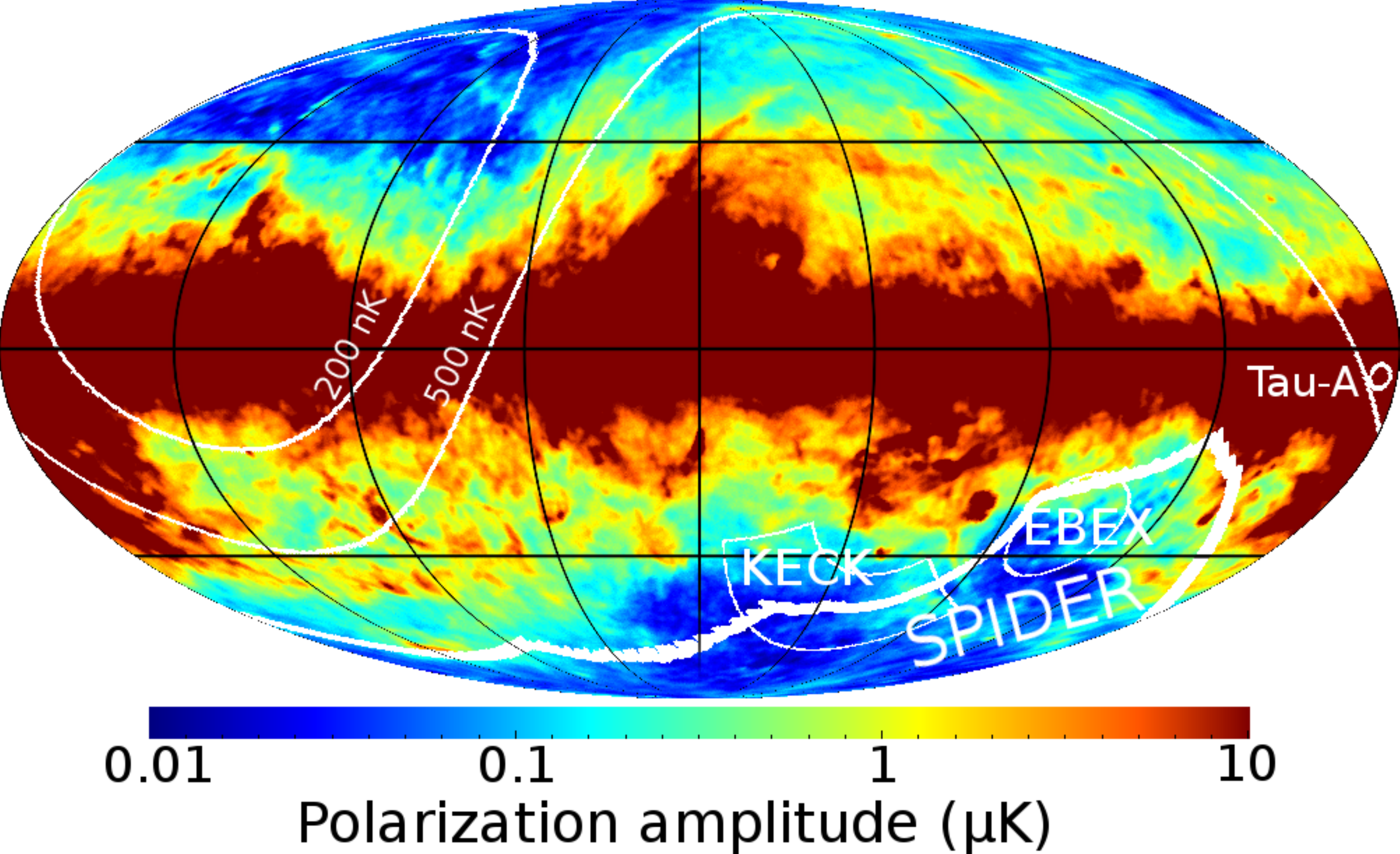}
   \end{tabular}
   \end{center}
   \caption[scanning] 
%>>>> use \label inside caption to get Fig. number with \ref{}
   { \label{fig:scanning} 
\textbf{Left:} The scan path contemplated by the feasibility study, projected on the sky in Galactic coordinates for a period of 10 days with a rotation speed artificially reduced to 0.01 RPM for clarity.  Without such reduction, the pattern appears completely dense and smooth.  \textbf{Right:} Comparison\cite{finkbeiner} of the observation field (upper left quadrant) with the SPIDER, BICEP/KECK and EBEX fields. The observation site and observing strategy are ideal for observing large-scale CMB polarization: No other B-mode experiment targets the Northern Hemisphere at frequencies $<$ 200 GHz, which Planck observes to be significantly cleaner than the Southern hemisphere; and the combination of a high --- but not too high --- observing latitude and fast detector read-out capabilities results in a scanning strategy with unprecedented cross-linking properties.}
\end{figure}

%%%%%%%%%%%%%%%%%%%%%%%%%%%%%%%%%%%%%%%%%%%%%%%%%%%%%%%%%%%%

\section{Observations}
\label{sec:observations}

The feasibility study we conducted used an example instrument configuration to highlight to utility of the Isi Station site in Greenland, which is currently being developed by the National Science Foundation.   The site lies at 72$^{\circ}$N, 38$^{\circ}$W and an altitude of 3,210 m.  The observing conditions at the site are excellent, with a precipitable water vapor (PWV) level roughly comparable to that of the well-known Chajnantor plateau in the Atacama desert at 23$^{\circ}$S, 68$^{\circ}$W and an altitude of 5,140 m~\cite{icecaps}.

Observing from the Greenland site would provide two significant advantages.  First, according to the Galactic thermal dust emission intensity and polarization fraction models recently released by Planck, the Northern Galactic hemisphere is appreciably cleaner with respect to thermal dust than the Southern, perhaps by as much as a factor of 2\cite{planckVI,planckII}. This translates into a foreground floor that is up to four times lower in the North than in the South, as measured in terms of B-mode power and tensor-to-scalar ratio.  Second, observing from a latitude of 72$^{\circ}$ would enable a uniquely cross-linked scanning strategy that mitigates a wide range of systematic errors, including correlated noise, beam asymmetries, intensity-to-polarization leakage, and calibration uncertainties. This scanning strategy is implemented by a telescope spinning continuously at a rate of 2$^{\circ}$ per second on the sky in segments of 24 hours each. Telescope elevation varies between 30 and 75$^{\circ}$ among segments, but is fixed within each segment to suppress atmospheric fluctuations. The resulting scan path is shown in Fig.~\ref{fig:scanning} Left. 
Notably, most pixels are observed for a wide range of directions, with the exception of those that lie at the outer edge of the field.
Furthermore, they are connected with pixels on the opposite side of the field on all time scales between $\sim$ 45 seconds and several months. This fast scanning strategy would not be feasible with slower detectors; it requires both a very fast read-out rate and a short detector time constant to avoid under-sampling.

%%%%%%%%%%%%%%%%%%%%%%%%%%%%%%%%%%%%%%%%%%%%%%%%%%%%%%%%%%%%%

\bibliography{article}
%\bibliography{report}   %>>>> bibliography data in report.bib
\bibliographystyle{spiebib}   %>>>> makes bibtex use spiebib.bst

%%%%%%%%%%%%%%%%%%%%%%%%%%%%%%%%%%%%%%%%%%%%%%%%%%%%%%%%%%%%%

\end{document}